\documentstyle{llncs}
\newcommand{\cancel}[1]{}

{\protect\end{tabbing}}
\begin{document}

\title{New Trends in Quantum Computing}

\author{Gilles {\sc Brassard}\,%
\thanks{Research supported in part by Canada's {\sc Nserc} and
Qu\'ebec's {\sc Fcar}.}}

\institute{
\mbox{Universit\'e de Montr\'eal},
\mbox{D\'epartement d'informatique et de recherche op\'erationnelle}\\
\mbox{C.\,P. 6128, Succursale Centre\,--Ville},
\mbox{Montr\'eal (Qu\'ebec)},
\mbox{{\sc Canada} H3C 3J7}\\
\mbox{email: \tt brassard@iro.umontreal.ca}}

\date{1 December 1995}

\maketitle

\begin{abstract}

Classical and quantum information are very different.
\mbox{Together} they can perform feats that neither could achieve alone,
such as quantum computing, quantum cryptography and quantum teleportation. 
Some of the applications range from helping to preventing spies from reading
private communications.  Among the tools that will facilitate their
implementation, we~note quantum purification and quantum error correction. 
\mbox{Although} some of these ideas are still beyond the grasp of current
technology, quantum cryptography has been implemented and the prospects are
encouraging for small-scale prototypes of quantum computation devices before
the end of the millennium.
\end{abstract}
\setcounter{footnote}{0}  

\vspace{-10cm}
{\raggedleft\small
{\it 13th Symposium on Theoretical Aspects of Computer Science}%
\hspace{-2cm}\mbox{}\\
\sl Grenoble, France 22\,--24 February 1996\hspace{-2cm}\mbox{}\\
\phantom{LNCS 1000,
Springer-$\!$Verlag, 1995}\hspace{-2cm}\mbox{}\\
\mbox{}}
\vspace{7.75cm}
\section{Introduction}

Classical and quantum information are very different.  \mbox{Classical}
information can be read, copied and transcribed into any medium; it can
be transmitted and broadcast.
Quantum information cannot be read or copied without disturbance,
but it can exist in superposition of classical states.
Together, the~two kinds of information
can perform feats that neither could achieve alone,
such as quantum computing, quantum cryptography and
quantum teleportation.  These concepts could result in a
revolution in computer science that may dwarf
that created decades ago by the transistor.

In~principle,
computers could be built to profit from quantum phenomena that have no
classical analogue, sometimes providing exponential speed-up compared to
classical computers.  The~most famous example of an algorithm
for the quantum computer, due to Shor, allows for the
polynomial-time factorization of large integers, a task believed to be
intractable for classical computers.  \mbox{Because} of the pivotal nature
of this problem in modern cryptography, a full-scale working
quantum computer could be used by spies to allow them nearly unlimited
access to your electronic transactions.  On~the other hand, quantum
information is also at the core of other phenomena that would be impossible to
achieve in a purely classical world, such as the unconditionally secure
distribution of secret cryptographic key material.  Therefore quantum
techniques may cause the collapse of much of classical cryptography, yet they
may also offer the cure to make unconditionally secure communication possible.

\section{Review of Quantum Techniques}

For a comprehensive review of quantum techniques in computer science,
I~suggest you read my earlier essay ``A~Quantum Jump in Computer
Science''~\cite{lncsmille}.
\mbox{Although} this section provides a very superficial one-page
introduction to these topics, my purpose in these Proceedings is
to report on new ideas and developments that were not yet available
at the time ``A~Quantum Jump'' went to press.
These developments range from new theoretical ideas to
actual implementation proposals for quantum computation.
To~emphasize how active and exciting the field has become,
I~deliberately restricted my coverage to papers that have appeared
in~1995 or later, or yet-unpublished manuscripts that
were written in the same period.  Please consult ``A~Quantum Jump''
for proper credit and references to the early ideas reviewed
in this section.

In~classical information theory, a~bit can take either value~0 or~1.
According to quantum mechanics, a~quantum bit,
or~{\em qubit}, can be in linear
\mbox{\em superposition\/} of the two classical states,
with complex coefficients.
This~is best \mbox{visualized} as a point on the surface of a unit sphere
whose North and South poles correspond to the classical values.
This is entirely different from taking a value {\em between\/} 0 and~1
as in classical analogue computing.
In~general, qubits cannot be measured reliably: not more than one
classical bit of information can be extracted from any given qubit
and the more information you obtain about~it, the more you disturb it
irreversibly.

The impossibility to measure quantum information \mbox{reliably} is at the
core of quantum cryptography.
When information is encoded with non-orthogonal quantum states, any
\mbox{attempt} from an eavesdropper to access it necessarily entails a
probability of spoiling it irreversibly, which can be detected by the
legitimate users.
This phenomenon can be exploited to implement a key distribution system that
is secure even against an eavesdropper with unlimited computing power.
Several prototypes have been built, including one that
is fully operational under laboratory conditions over 30~kilometres of
ordinary optical fibre.  In~another experiment, quantum transmission
was successful over a distance of 23~kilometers under Lake Geneva~\cite{geneva}.
Quantum techniques may also assist in the achievement of subtler
cryptographic goals, such as protecting private information while it is
being used to reach public decisions.

Independent qubits are sufficient to produce nontrivial cryptographic
phenomena, but they are not very interesting for computational purposes.
For~this, we must consider quantum {\em registers\/}
composed of $n$ qubits.  Such registers can be in an
arbitrary superposition of all $2^n$ classical states.
In~principle, a quantum computer can be programmed so that exponentially many
computation paths are taken simultaneously in a single piece
of hardware, a~phenomenon known as {\em quantum parallelism}.
What makes this so powerful---and~mysterious---is
the exploitation of constructive and destructive {\em interference}, which
allows for the reinforcement of the probability of obtaining desired results
while at the same time the probability of spurious results is reduced or even
annihilated.  In the words of Feynman,
``it appears as if the probabilities would have to go negative''.

\section{Good Starting Points}

In addition to my ``Quantum Jump'' article~\cite{lncsmille},
several excellent introductions to quantum computing have
been written recently.  Let~us mention
\cite{lloyd:sciam} in {\it \mbox{Scientific} American},
\cite{discover} in {\it Discover},
\cite{engineering} in {\it Nature},
\cite{physics:today} in {\it Physics Today},
\cite{div:science} in \mbox{\it Science\/},
and~\cite{CTR}.
In~addition, Shor has written a very nice account of
his quantum factorization algorithm~\cite{shor:improved}, which
introduces the basic concepts of quantum computing.

\section{The Power of Quantum Computing}

Quantum computing was considered at best as a curiosity by most researchers
until Shor discovered in 1994 that it could be used to extract
discrete logarithms and factorize large numbers very efficiently.  This
attracted considerable attention, not only because of its tremendous
cryptographic significance, but also because it gave the first indication that
quantum computers could be genuinely faster than classical probabilistic
computers for solving natural problems of a mathematical nature.  The~obvious
question that followed was: ``What else are quantum computers good~at?''

Unfortunately, Shor's discovery was not followed by a flurry of other natural
problems that the quantum computer could solve much more efficiently than
using the best algorithm available for a classical computer.  Nevertheless,
a~few such results have emerged already and perhaps still others are waiting
in the wings.  Using a method similar to Shor's, Boneh and Lipton~\cite{lipton}
showed that any cryptosystem based on what they call a ``hidden linear form''
can be broken in polynomial time on a quantum computer.  In~particular, a
quantum computer can solve the discrete logarithm problem efficiently over
{\em any\/} group including Galois fields and elliptic curves.

Another extension to Shor's result is due to
Kitaev~\cite{kitaev}, who discovered how to solve the Abelian stabilizer
problem efficiently on a quantum computer.  This~method provides an efficient
quantum Fourier transform algorithm for an arbitrary Abelian group.

\section{Experiments in Quantum Computing}

As~I explained in ``A~Quantum Jump'', the discovery that universal quantum
circuits could be built around two-qubit gates was very significant because
the technological difficulties would be even more daunting if it had been
necessary to make qubits interact three at a time.  Unfortunately I~gave
only reference to the work of DiVincenzo~\cite{divin:two} in my earlier
essay.
The~same result was found independently by Barenco~\cite{universal:barenco}
and by Sleator and Weinfurter~\cite{sleator}.
It~was subsequently discovered by Chau and Wilczek~\cite{fredkin} that six
two-qubit gates are sufficient to implement the quantum Fredkin gate,
which is a natural three-qubit universal gate; this result was improved
by Smolin and DiVincenzo~\cite{smo:div} to needing only five two-qubit gates
for the same purpose.  In~fact, almost {\em any\/} two-qubit gate is universal,
as Deutsch, Barenco and Ekert~\cite{universality} and Lloyd~\cite{lloyd}
discovered independently.  The most significant result along these lines
is probably the discovery that the quantum exclusive-or gate,
which maps \mbox{$(x,y)$} to \mbox{$(x,x \oplus y)$}, is universal for
quantum computation in the sense that all unitary operations on arbitrarily
many qubits can be expressed as compositions of these gates together with
appropriate one-qubit gates~\cite{gangofnine}.

This begs the question: how hard is it to implement the quantum
exclusive-or gate?  Several approaches have been proposed for this
purpose.  Cirac and Zoller proposed to use cold trapped
ions~\cite{cirac:zoller}.  An~actual implementation of the quantum
exclusive-or gate using beryllium ions in an atom trap has been
tested with encouraging results by Monroe, Meekhof, King, Itano
and Wineland~\cite{monroe} at NIST in Boulder, Colorado.
Another team led by Hughes at the Los~Alamos National Laboratory
has received funding for experimenting with cold trapped calcium ions.
They hope to be able to implement a small-scale version of Shor's
quantum factorization algorithm within a few year.  Their initial goal
is to factorize the number~4 before the end of the millennium, but they are
confident that this will only be a beginning.

Another approach to the implementation of basic quantum gates, using
cavity quantum electrodynamics, has been proposed by Sleator and
Weinfurter~\cite{sleator}.  Initial experiments on similar ideas,
using atomic interferometry and microwave cavities, have been performed
at the \'Ecole Normale Sup\'erieure in Paris by Domokos, Raimond,
Brune and Haroche~\cite{haroche}.  Experiments are also under way
at the California Institute of Technology, where Turchette, Hood,
Lange, Mabuchi and Kimble are investigating photon qubits interacting in an
optical microcavity~\cite{turchette}.
See~also~\cite{BDEJ}.

An explicit construction of quantum networks for performing arithmetic
operations from basic quantum gates is given by Vedral, Barenco and
Ekert~\cite{VBE}.  This may prove important for an actual implementation
of Shor's algorithm, which requires basic arithmetic from addition
to modular exponentiation.  In~particular, this paper shows that the
amount of auxiliary storage required to implement Shor's algorithm
grows linearly with the size of the number to be factorized.

Yet another proposal for the construction of a ``simple quantum
computer'' comes from Chuang and Yamamoto~\cite{yamamoto}.

\section{The Problem of Decoherence}

Despite the reasons to be optimistic that the work described in the
preceding section might generate, it may be that
quantum computing will never become practical because of the
technological difficulties in preventing unwanted interactions
with the environment: such interactions cause {\em decoherence},
which in effect ruins the quantum computation.  An~early---and~rather
discouraging---study of the effect of decoherence on quantum computers
was carried out by Unruh~\cite{unruh}.  Other difficulties
with the implementation of quantum computers
have also been pointed out repeatedly by Landauer~\cite{landauer}.

The effect of decoherence on Shor's algorithm
has been studied explicitly by Chuang, Laflamme, Shor and Zurek~\cite{CLSZ}
and by Plenio and Knight~\cite{knight}.

\section{Quantum Error Correction}

Error correction is used routinely when dealing with classical information. 
\mbox{However}, it is not obvious that error-correction schemes can exist for quantum
information because it cannot be measured without disturbance.  In~particular,
a simple repeat code is out of the question since quantum information cannot be
cloned.  Nevertheless, quantum information needs to be protected from errors
even more than classical information in view of its susceptibility to
decoherence.  An~early scheme for quantum error correction, proposed by
Deutsch, was investigated by Berthiaume~\cite{andre} and Jozsa.

Assuming that the decoherence process affects the quantum computer's
qubits independently, Shor has found a technique that allows the storage
of an arbitrary state of $n$ qubits using $9n$ qubits in a
decoherence-resistant way: even if one of the qubits decoheres, the original
state can be reconstructed perfectly~\cite{shor:coding}.
Subsequently, Shor improved on his original idea in collaboration with
Calderbank, making it possible to recover the original quantum
state even if several qubits decohere~\cite{calderbank}.
Other quantum error-correction
techniques have been proposed by Chuang and Laflamme~\cite{laflamme}
and by Steane~\cite{steane}.

Another approach to quantum error correction is based on the ideas
of \mbox{entanglement} concentration~\cite{concentration}
and entanglement purification~\cite{BBPSSW}.
The~latter is a technique that allows near perfect entanglement
to be distilled from imperfect entanglement that may have been caused
by partial decoherence---or~by eavesdropping for quantum cryptographic
applications.  This is accomplished by local operations and the exchange
of classical messages.  Because perfect entanglement can be used for
the purpose of teleporting quantum information, entanglement purification
can be used to transmit quantum information with arbitrary fidelity
over a noisy quantum channel supplemented by a good classical channel.
More advanced ideas, such as ``teleportation from the present to the
future, rather than from here to there'' can be used to design
a quantum error-correction scheme from quantum purification and
quantum teleportation~\cite{BDSW}.

\section{The Art of Quantum Programming}

By now, I am sure you are itching to write your first program for
the quantum computer.  In~that case, you will be happy to learn that
Baker is in the process of developing {\sc Q--Gol}, a~high-level
language for the quantum computer.   You~can find more information
on the World Wide Web (WWW) at URL
\begin{quote}
 \verb+http://www.rp.csiro.au/~gbaker/q-gol/+
\end{quote}

\section{Quantum Information on the Internet}

Many WWW sites have blossomed with information on quantum computation,
quantum cryptography and quantum information theory in general.
The~following URLs are excellent starting points for
a fascinating journey into the quantum world.
Nearly all the papers cited as ``manuscript'' or ``in~press''
in the references at the end of this essay
can be downloaded from the quantum physics archive at
Los~Alamos National Laboratory.  Have~fun!

\begin{quote}
\verb+http://xxx.lanl.gov/archive/quant-ph+\\
\hspace*{1cm}Quantum Physics Archive at Los Alamos National Laboratory\\[1mm]
\verb+http://eve.physics.ox.ac.uk/QChome.html+\\
\hspace*{1cm}Quantum Computation and Cryptography Home Page at Oxford\\[1mm]
\verb+http://aerodec.anu.edu.au/~qc/index.html+\\
\hspace*{1cm}Quantum Computing Home Page\\
\hspace*{1cm}at Australian National University\\[1mm]
\verb+http://feynman.stanford.edu/qcomp/+\\
\hspace*{1cm}Quantum Computation Archive at Stanford\\[1mm]
\verb+http://vesta.physics.ucla.edu/~smolin/+\\
\hspace*{1cm}John Smolin's Quantum Information Page\\[1mm]
\verb+http://www.cwi.nl/~berthiau+\\
\hspace*{1cm}Andr\'e Berthiaume's Home Page\\[1mm]
\verb+http://www.iro.umontreal.ca/labs/theorique/index_en.html+\\
\hspace*{1cm}Laboratoire d'informatique th\'eorique et quantique\\
\hspace*{1cm}at Universit\'e de Montr\'eal
\end{quote}

\noindent
In addition, you can find tutorials at the following URLs.

\begin{quote}
\verb+http://chemphys.weizmann.ac.il/~schmuel/comp/comp.html+\\
\hspace*{1cm}Samuel L. Braunstein's Tutorial on Quantum Computation\\[1mm]
\verb+http://eve.physics.ox.ac.uk/QCresearch/cryptoanalysis/qc.html+\\
\hspace*{1cm}Artur Ekert's ``Introduction to Quantum Cryptoanalysis''\\[1mm]
\verb+http://www.cwi.nl/~berthiau/publications/CTR.ps+\\
\hspace*{1cm}Andr\'e Berthiaume's Tutorial on Quantum Computation~\cite{CTR}
\end{quote}

\section{Do you Need a Daily Fix?}

If you cannot live without knowing what is new every day,
send electronic mail to {\tt quant-ph@xxx.lanl.gov}
with subject ``subscribe'' and let it guide you!

\section*{Acknowledgements}

I am very grateful to all those who helped me put this essay together:
Charles H. Bennett,
David DiVincenzo,
Artur Ekert,
Christopher Fuchs,
Richard Hughes,
Michael Kernaghan,
Christopher Monroe,
Peter Shor and
Andrew Steane.

\end{document}